\documentclass[10pt,twocolumn,showpacs,longbibliography,prl,aps,floatfix]{revtex4-1}

\usepackage{graphicx}
\usepackage{dcolumn}
\usepackage{float}
\usepackage{amsmath}
\usepackage{amssymb}
\usepackage{bm}

\newcommand{\comment}[1]{}
\newcommand{\eps}{\varepsilon}

\begin{document}

\title{Magneto-optics of massless Kane fermions: Role of the flat band and unusual Berry phase}

\author{J.D. Malcolm}
\author{E.J. Nicol}
\affiliation{Department of Physics, University of Guelph, Guelph, Ontario N1G 2W1 Canada} 
\affiliation{Guelph-Waterloo Physics Institute, University of Guelph, Guelph, Ontario N1G 2W1 Canada}
\affiliation{Kavli Institute for Theoretical Physics, University of California, Santa Barbara, CA 93106 USA}
\date{\today}

\begin{abstract}
{${\rm Hg}_{1-x}{\rm Cd}_{x}{\rm Te}$ at a critical doping $x=x_c\approx 0.17$ has a bulk dispersion which includes two linear cones meeting at a single point at zero energy, intersecting a nearly flat band, similar to the pseudospin-1 Dirac-Weyl system.  In the presence of a finite magnetic field, these bands condense into highly degenerate Landau levels.  We have numerically calculated the frequency-dependent magneto-optical and zero-field conductivity of this material using the Kane model.  These calculations show good agreement with recent experimental measurements.  We discuss the signature of the flat band and the split peaks of the magneto-optics in terms of general pseudospin-$s$ models and propose that the system exhibits a non-$\pi$-quantized Berry phase, found in recent theoretical work.
}
\end{abstract}



\maketitle

\noindent\emph{Introduction}.  With the development of condensed matter Dirac systems, much research has focused on flat bands and non-trivial Berry phases, among other features.  The macroscopic degeneracy found in dispersionless, or flat, bands produces a singular density of states, potentially opening the door to some interesting physics where interactions can lift this degeneracy.  In the presence of a magnetic field, highly degenerate Landau levels (LL's) are formed out of continuous-dispersion systems.  At partial filling of these levels, interactions between electrons give rise to the fractional quantum Hall effect~\cite{Tsui82,Laughlin83}.  In addition, room-temperature superconductivity has been proposed in discussions of flat bands present on the surfaces of topological media~\cite{Heikkila11}.  Another feature of many Dirac materials is the non-trivial Berry phase.  Such gives rise to both the half-integer Hall conductivity and magneto-oscillation shift seen in graphene, for example~\cite{Zhang05,CastroNeto09}.  Most recently, a variable Berry phase model has been proposed which theoretically tunes the magnetic response of a Dirac system from diamagnetic to paramagnetic~\cite{Raoux14}.

In contemporary literature, ${\rm Hg}_{1-x}{\rm Cd}_x{\rm Te}$ (MCT) is typically discussed in the context of quantum wells and the quantum spin-Hall effect~\cite{Bernevig06,Konig07}.  However, a particular phase of the bulk material that exhibits a nominally flat heavy-hole band at zero energy is also quite exciting in its similarity to Dirac materials~\cite{Orlita14}.  This phase exists at critical cadmium concentration $x = x_c \approx 0.17$, marking the transition between distinct phases: semimetal for $x<x_c$ and semiconductor for $x>x_c$.  The flat band provides its own signature in the magneto-optical response of the material, much like in Dirac-Weyl systems~\cite{Malcolm14}.  Within this paper, we provide a numerical calculation of the bulk optical conductivity for MCT, showing complete spectral-weight dependence on photon frequency both in the presence and absence of a magnetic field.  This allows for direct comparison to a recent experimental measurement and analysis of MCT's optical properties by Orlita \emph{et al}.~\cite{Orlita14}.  We are able to show excellent agreement between theory and experiment and provide further insight into the signature and role of the flat band in this material.  Moreover, we show that this system can be linked to the $\alpha$-$T_3$ model \cite{Raoux14} which has non-$\pi$-quantized Berry phase.\newline

\noindent\emph{Kane Model}.  MCT at critical concentration $x_c$ is described by a reduced Kane model Hamiltonian \cite{Kane57,Orlita14},
\begingroup
\renewcommand*{\arraystretch}{1.5}
\begin{widetext}
\begin{equation}\label{eqn:KaneHam}
\hat{\mathcal{H}}_K = \hbar v
\begin{pmatrix}
0 & \frac{\sqrt{3}k_-}{2} & 0 & 0 & 0 & 0 & 0 & 0 \\
\frac{\sqrt{3}k_+}{2} & \frac{E_g}{\hbar v} & -\frac{k_-}{2} & -\frac{k_-}{\sqrt{2}} & -\frac{k_z}{\sqrt{2}} & -k_z & 0 & 0 \\
0 & -\frac{k_+}{2} & 0 & 0 & 0 & 0 & -k_z & 0 \\
0 & -\frac{k_+}{\sqrt{2}} & 0 & -\frac{\Delta}{\hbar v} & 0 & 0 & -\frac{k_z}{\sqrt{2}} & 0 \\
0 & -\frac{k_z}{\sqrt{2}} & 0 & 0 & -\frac{\Delta}{\hbar v} & 0 & -\frac{k_-}{\sqrt{2}} & 0 \\
0 & -k_z & 0 & 0 & 0 & 0 & \frac{k_-}{2} & 0 \\
0 & 0 & -k_z & -\frac{k_z}{\sqrt{2}} & -\frac{k_+}{\sqrt{2}} & \frac{k_+}{2} & \frac{E_g}{\hbar v} & -\frac{\sqrt{3}k_-}{2} \\
0 & 0 & 0 & 0 & 0 & 0 & -\frac{\sqrt{3}k_+}{2} & 0 \\
\end{pmatrix},
\end{equation}
\end{widetext}
\endgroup
\noindent whose parameters include $v$, a velocity characteristic to the material; $E_g$, a small energy gap; $\Delta$, the spin-orbit splitting providing a large band separation; and where $k_{\pm} = k_x \pm ik_y$.  This model is only first order in momentum, which approximates the broad curvature in the heavy hole bands of the actual material as being flat.  The form of the Kane Hamiltonian in Eq.~(\ref{eqn:KaneHam}) is obtained from a previous presentation~\cite{Orlita14} through a simple permutation of the basis states.  Note, the limit $\Delta\rightarrow\infty$ decouples the fourth and fifth columns from the others, giving an effective $6\times 6$ model which for $E_g=k_z=0$ maps to a model with an unusual Berry phase (discussed below).  The presence of a finite nonzero $\Delta$ acts to break particle-hole symmetry.

\begin{figure}
\begin{center}
\includegraphics[width=1.0\linewidth]{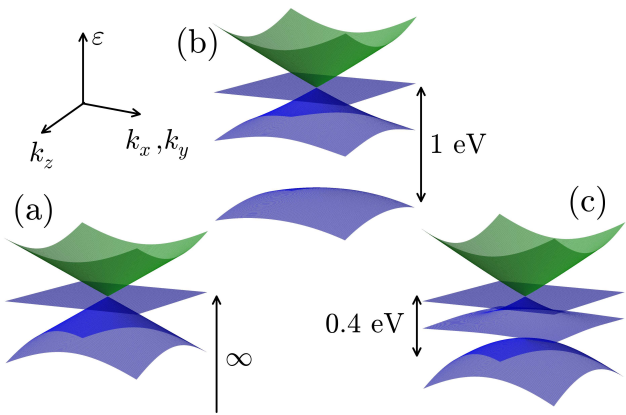}
\end{center}
\caption{\label{fig:Dispersion}(Color online) Kane fermion dispersion for zero magnetic field with the parameters $v = 1.06\times10^6\,{\rm m/s}$, $E_g = 4\,{\rm meV}$, and (a) $\Delta \rightarrow \infty$, (b) $\Delta = 1\,{\rm eV}$, (c) $\Delta = 0.4\,{\rm eV}$.}
\end{figure}

Using the parameters of $v=1.06\times10^6{\rm m/s}$ and $E_g=4\,{\rm meV}$ taken from Ref.~\cite{Orlita14}, the so-called Kane fermion dispersion is shown in Fig. \ref{fig:Dispersion} for different values $\Delta = 0.4\,{\rm eV}$, $\Delta = 1\,{\rm eV}$, and the limit $\Delta \rightarrow \infty$. Each band in the figure is doubly degenerate and the upper/lower green/purple bands are unoccupied/occupied.  We see that for the infinite separation value in $\Delta$, the dispersion resembles the Weyl system with pseudospin $s=1$~\cite{Dora11,Malcolm14}, although Eq.~(\ref{eqn:KaneHam}) does not map exactly onto this Hamiltonian.  When in close vicinity, the bottom band distorts the lower cone away from linearity, while narrowing the upper cone, seen in the progression between panels (a)-(c).  For all subsequent calculations in modelling MCT, the value of $\Delta = 1\,{\rm eV}$ was used.\newline

\noindent\emph{Zero-Field Optics}.  Using the general Hamiltonian in Eq.~(\ref{eqn:KaneHam}), we can calculate the zero-field conductivity at different photon energy, $\Omega$, via the Kubo formula~\cite{Mahan81},
\begin{equation}\label{eqn:cond-zero-field}
\begin{split}
&{\rm Re}\,\sigma_{xx}(\Omega) \\ &=\frac{\hbar e^2}{8\pi^2} \sum_{\lambda,\lambda'}\int d^3\bm{k}\frac{\Delta n_f}{\Delta\eps} \left|\left\langle\lambda'\right|\hat{v}_x\left|\lambda\right\rangle\right|^2 \mathfrak{L}(\Omega-\Delta\eps,\eta),
\end{split}
\end{equation}
where the summation is over transitions from a state in the initial band $\lambda$ with energy $\eps$ to a final state in band $\lambda'$ of energy $\eps'$.  $\hat{v}_x=\partial\hat{\mathcal{H}}/\partial(\hbar k_x)$ is the velocity operator and $\mathfrak{L}(x,\eta)=\eta/[\pi(x^2+\eta^2)]$ is a Lorentzian function centred at $x=0$ with a full width at half maximum of $\eta$, the scattering rate, taken to be $2\,{\rm meV}$.  $\Delta\eps = \eps' - \eps$ and $\Delta n_f = n_f(\eps) - n_f(\eps')$, where $n_f$ is the Fermi-Dirac distribution at chemical potential $\mu = 0^+$, which ensures a filled flat band.

\begin{figure}
\begin{center}
\includegraphics[width=1.0\linewidth]{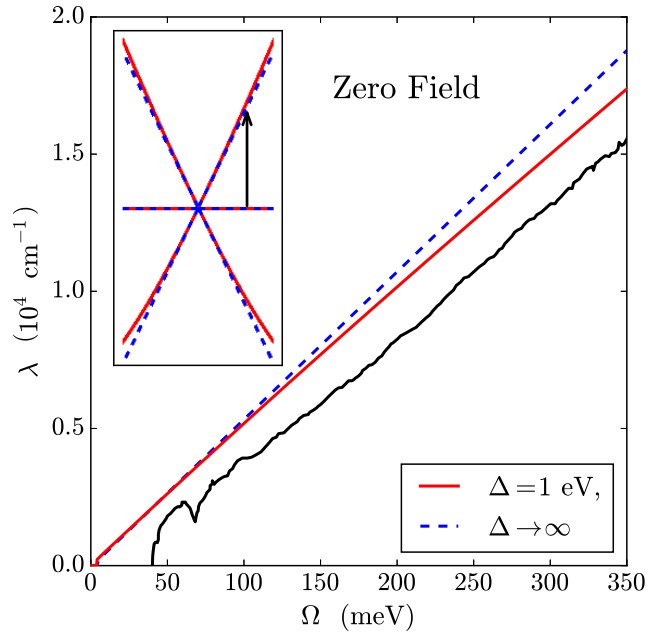}
\end{center}
\caption{\label{fig:ZeroField} (Color online) Kane fermion absorption coefficient, $\lambda$, for different values of parameter $\Delta$ (solid red and dashed blue) plotted against the zero-field experimental measurement taken from Ref.~\cite{Orlita14} (solid black line).  Inset: cross section of relevant band structures, showing asymmetry in the red curve, with a typical transition from the flat band.}
\end{figure}

The red (solid) line in Fig.~(\ref{fig:ZeroField}) is the result of a numerical calculation of Eq.~(\ref{eqn:cond-zero-field}).  This is plotted for comparison with the MCT absorption coefficient, $\lambda=\sqrt{4\Omega\sigma/\epsilon_0\hbar c^2}$, measured experimentally as the black (solid) line.  Note that the experimental data are cut off below around $40$ meV by the Restrahlen band (see Ref.~\cite{Orlita14}) and omission of a low-frequency phonon peak.  The blue (dotted) line is $\lambda$ in the approximation $\Delta\rightarrow\infty$~\cite{Orlita14}.  The major component of the spectral weight in the calculated $\lambda$ is due to flat-to-cone transitions between bands.  Linear behaviour is exhibited in both the Kane model results and the physical MCT measurement, akin to the 3D Weyl system \cite{Ashby13} discussed below.  In comparison, we see that the red theoretical curve for $\Delta = 1\,{\rm eV}$ provides a better match to the slope in the experimental curve, but the theory remains offset above the data.  The non-zero intercept extrapolated from the experimental data may have arisen from a small unaccounted-for mismatch in the dielectrics of the MCT and its substrate \cite{Orlita14}.  Linear conductivity is seen in some quasicrystal optical responses as well, where a negative vertical intercept there has been possibly attributed to an unusual gapped Dirac point~\cite{Timusk13}.  The better match using finite $\Delta$ demonstrates the importance of particle-hole asymmetry whereby the upper cone is narrowed (see inset), reducing the associated density of states and absorption. \newline

\noindent\emph{Magneto-Optics}.  At the introduction of a magnetic field $\bm{B}=\nabla\times\bm{A}=B\hat{e}_z$, a Peierls substitution is made in the momentum, $\bm{k}\rightarrow \bm{k}+e\bm{A}/\hbar c$.  This allows one to rewrite the Hamiltonian in terms of ladder operators, $k_+\rightarrow \sqrt{2}a^\dagger/\ell_B,\,k_-\rightarrow \sqrt{2}a/\ell_B$.  $\ell_B=\sqrt{\hbar/e|B|}$ is the magnetic length scale.  The operators act on Fock degrees of freedom, $\left|m\right\rangle$, found in the energy eigenvector, with $a\left|m\right\rangle=\sqrt{m}\left|m-1\right\rangle$, $a^\dagger\left|m\right\rangle=\sqrt{m+1}\left|m+1\right\rangle$, and $[a,a^\dagger]=1$.  The wavefunction for each LL, $\left|\psi^\lambda_n\right\rangle$, gets labelled with a Fock number $n$ and a band index $\lambda$.  
In this finite-field case, we can make use of the 3D Kubo formula written now in the LL basis,
 \begin{equation}\label{eqn:cond-LL-basis}
\begin{split}
&{\rm Re}\,\sigma_{xx}(\Omega) \\
& = \frac{\hbar e^2}{4\pi\ell_B^2} \sum_{\psi,\psi'}\int_{-\infty}^\infty dk_z\frac{\Delta n_f}{\Delta\eps} \left|\left\langle\psi'\right|\hat{v}_x\left|\psi\right\rangle\right|^2\mathfrak{L}(\Omega-\Delta\eps,\eta).
\end{split}
\end{equation}
The summation on $\psi$ in Eq. (\ref{eqn:cond-LL-basis}) is taken over band index $\lambda$ and Fock number $n$.

\begin{figure}
\begin{center}
\includegraphics[width=1.0\linewidth]{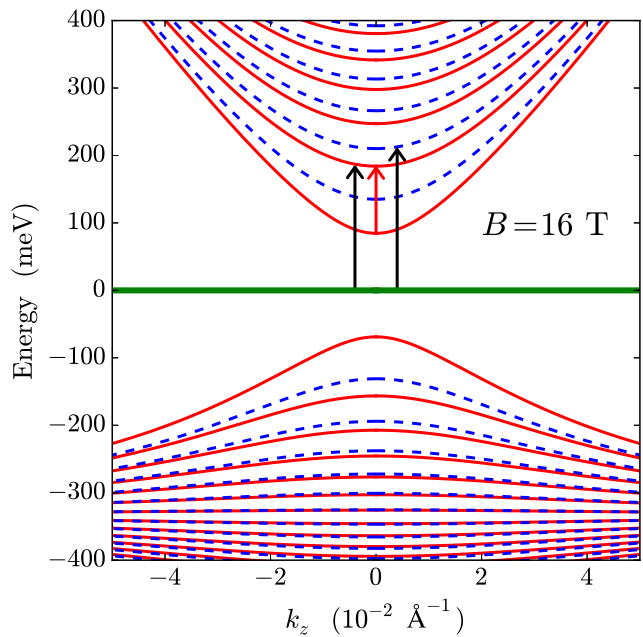}
\end{center}
\caption{\label{fig:LandauLevels}(Color online) Dispersive Landau levels in the 3D Kane system under a $16\,{\rm T}$ magnetic field with $E_g=4\,{\rm meV}$ and $\Delta = 1\,{\rm eV}$.  At $k_z=0$, red (solid) bands reside in Sector A and the blue (dashed) bands in Sector B .  At zero energy (green) there are many Landau bands which are in either sector at $k_z=0$.  Illustrated are transitions that are responsible for the peaks indicated in the next figure.}
\end{figure}

\begin{figure}
\begin{center}
\includegraphics[width=1.0\linewidth]{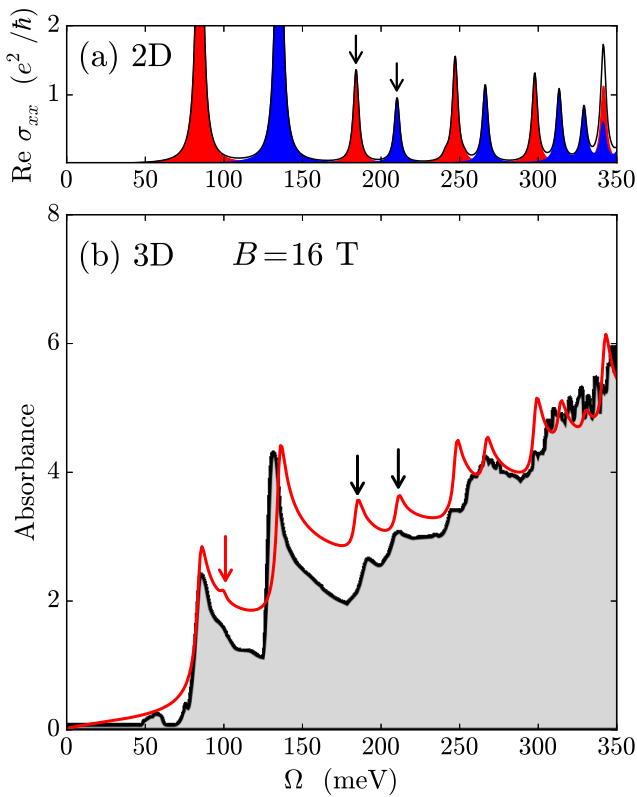}
\end{center}
\caption{\label{fig:BField}(Color online) (a) Magneto-optics of the 2D massless Kane fermion system.  Red shading indicates contributions from transitions in Sector A, while blue those from Sector B.  The sum of these is the total conductivity in black. (b) Magneto-optical absorbance for the 3D system in a $16\,{\rm T}$ magnetic field in red (solid).  In the background is the MCT absorbance measurement in black (solid) from Ref.~\cite{Orlita14}.  Indicated in both plots are the secondary peaks straddling $200\,{\rm meV}$ found in each Weyl-like spectrum and discussed in the text.  In (b), the red arrow indicates the cyclotron resonance peak.}
\end{figure}

With a finite magnetic field, the double degeneracy of the bands in Fig.~\ref{fig:Dispersion} is lifted as they condense into LL's that disperse along $k_z$ (Fig.~\ref{fig:LandauLevels}).  These bands carry a large density of states at each value of momentum $k_z$.  At the point $k_z=0$, the Hilbert space of Eq.~(\ref{eqn:KaneHam}) decomposes into two independent sectors, with the upper $4\times 4$ block being referred to as Sector A and the lower block Sector B.  In the simplified limit of $\Delta\rightarrow\infty$ and $E_g=0$, the 2D ($k_z=0$) Sector A provides LL's quantized with energies $\eps_{\rm 2D}^A=\gamma\sqrt{4n-7}$ ($n\geq 2$) in units of $\gamma=\hbar v/\sqrt{2}\ell_B$.  Sector B, however, allows levels with a different energy spectrum, $\eps_{\rm 2D}^B=\gamma\sqrt{4n-1}$ ($n\geq 1$).  In Fig.~\ref{fig:LandauLevels}, red (solid) bands belong to Sector A at $k_z=0$ and blue (dashed) bands to Sector B.  The green (solid) flat band at zero energy consists of many LL's that are in either sector at $k_z=0$.  Restricted to this 2D limit, optically activated transitions between the two sectors are strictly forbidden.  The result is an optical conductivity, made up of two congruous spectra from each sector, shifted in energy.  This was calculated using the 2D version of  Eq.~(\ref{eqn:cond-LL-basis}) for a magnetic field strength of $16\,{\rm T}$ and is shown in Fig.~\ref{fig:BField}(a).  The conductivity in Sector A (red spectrum) is shifted to lower energy relative to Sector B (blue), but shares the same form.  Each peak describes optically activated transitions between LL's at particular energies which obey the selection rule $n\rightarrow n\pm 1$.  The majority of features seen are due to excitations out of the flat band into the conduction band.  Excitations out of the lower cone begin to appear at higher energies and are relatively suppressed (see Ref.~\cite{Malcolm14}). For example, the small shoulder on the left of the red peak seen near $240\,{\rm meV}$ and the last two blue peaks all come from cone-to-cone transitions.  Referring to the flat-to-cone series of peaks, we see that the reduced height of the second peak in each sector (indicated by arrows) produces a \emph{non-monotonic} decline in the peak heights.  We have recently predicted this same effect also in the 2D Dirac-Weyl systems with integer pseudospin-$s$, where it indicates the presence of a flat band~\cite{Malcolm14}.  The particular signature in the Kane model of a single reduced peak points specifically to its pseudospin-1 nature.

In moving to the full 3D conductivity, the extra dimension of dispersion does not change the location of the 2D peaks, but merely adds a tail to them stretching out toward high energies.  Tails from neighboring peaks add together to build an overall linear profile, having been described in the context of the hypothetical 3D pseudospin-$1/2$ Weyl system by Ashby and Carbotte~\cite{Ashby13}.   This extension to 3D is seen in the result of Eq.~(\ref{eqn:cond-LL-basis}) presented in Fig.~\ref{fig:BField}(b) as absorbance, $A=d\lambda$, for $B=16\,{\rm T}$ and a sample thickness of $d=3.2\,{\rm \mu m}$.  In the background of the figure is the absorbance of MCT at $16\,{\rm T}$ for comparison~\cite{Orlita14}.  By slightly filling the first positive LL, we have been able to construct the cyclotron resonance peak in the quantum limit (red arrow in Figs.~\ref{fig:LandauLevels} and \ref{fig:BField}(b)), which is seen at the same energy in the experiment.  Also indicated are those secondary peaks from the flat band, which retain their reduced height characteristic.  The frequency-dependent conductivity calculated here offers a strong agreement between theory and experiment and we see that the measured absorbance too displays the secondary peaks with a reduced height (a sign of the pseudospin-1 dynamics in MCT).  This latter fact demonstrates the broad curvature of the MCT heavy hole band which can be sufficiently approximated as flat.  As in the zero-field calculation shown in Fig.~\ref{fig:ZeroField}, there is a vertical offset between the two data sets, the possible source of which is discussed in the preceding section.\newline

\noindent\emph{Weyl System}.  Throughout this paper, there has been reference to the pseudospin-1 nature of the Kane model, particularly seen in the individual Sectors A and B separately.  Note, however that each sector does not map to the purely $s=1$ Weyl Hamiltonian,
\begin{equation}
\hat{\mathcal{H}}_W^s = \hbar v\hat{\pmb{S}}\cdot\pmb{k},
\end{equation}
where $\hat{\pmb{S}}$ is the set of pseudospin-$s$ matrices.  Instead, in the limits $E_g=0$ and $\Delta\rightarrow\infty$, the 2D ($k_z=0$) Sectors A and B can be seen to be an admixture of $s=1$ and $s=1/2$ Weyl systems.  For instance, around Sector A, the $3\times3$ Hamiltonian is
\begin{equation}\label{eqn:SectorA}
\begin{split}
\hat{\mathcal{H}}_A &= \frac{\sqrt{2}\alpha}{\sqrt{1+\alpha^2}}\,\hat{\mathcal{H}}_W^1+\frac{2(1-\alpha)}{\sqrt{1+\alpha^2}}\,\left(\hat{\mathcal{H}}_W^{1/2}\oplus0\right)
\end{split}
\end{equation}
with $\alpha=1/\sqrt{3}$.  As a single matrix, one sees that Eq.~(\ref{eqn:SectorA}), up to a simple unitary transformation, describes the low-energy physics around the $K$ point in the $\alpha$-$T_3$ model proposed by Raoux et al. \cite{Raoux14},
\begin{equation}\label{eqn:alphaT3}
\begin{split}
\hat{\mathcal{H}}_\alpha
&= \frac{\hbar v}{\sqrt{1+\alpha^2}}
\begin{pmatrix}
0 & k_- & 0 \\
k_+ & 0 & \alpha k_- \\
0 & \alpha k_+ & 0
\end{pmatrix}.
\end{split}
\end{equation}
Similarly, Sector B maps to the $K'$ valley index in the Raoux \emph{et al}. model.  In Eqs.~(\ref{eqn:SectorA}) and~(\ref{eqn:alphaT3}), the value $\alpha=1$ corresponds to the $s=1$ Weyl system and $\alpha=0$ to the $s=1/2$ (graphene) system with a dormant uncoupled flat band.  Intermediate values of $\alpha$ are an admixture of both and exhibit bands with a non-$\pi$-quantized Berry phase.  For $\alpha=1/\sqrt{3}$ one determines that the Berry phases assigned to the lower, flat, and upper cones are $(\pi/2,-\pi,\pi/2)$, respectively.  Physical MCT, existing in three dimensions, couples the two valleys through a non-zero $k_z$, with additional corrections provided by $\Delta$ and $E_g$.  Raoux \emph{et al}. proposed an experimental realization of the $\alpha$-$T_3$ model in an optical lattice loaded with cold fermionic atoms.  With our new insight, we suggest that MCT could in addition provide a solid-state analogue for this model with a unique Berry phase, manifest in the relative shift of magneto-optical absorption peaks between sectors in Fig.~\ref{fig:BField} \cite{misc:Illes}.\newline

\noindent\emph{Summary}.  We have calculated the frequency-dependent magneto-optical response of the massless Kane fermion MCT system, providing a rigorous quantitative prediction of the optical spectral weight under each line.  Good agreement with experimental data was obtained by applying a reduced Kane model which approximates the material's heavy hole band to be exactly flat.  Moreover, we have been able to demonstrate the kinship that these Kane fermions possess with the appropriate Dirac-Weyl counterpart in the $\alpha$-$T_3$ model, which gives rise to a split-peak magneto-optical spectrum and points to an unusual Berry phase.  In addition, the cyclotron resonance peak in the quantum limit has been identified in both theory and experiment.  MCT continues to offer many opportunities for exploration here in the associated field of Dirac materials with unusual Berry phase.

We thank M. Orlita for detailed feedback on this work and pointing out the cyclotron resonance feature.  We also acknowledge J.P. Carbotte, E. Illes, B. Pavlovic, and C.J. Tabert for helpful discussions.  This work has been supported by the Natural Sciences and Engineering Research Council (NSERC) of Canada and by the National Science Foundation under Grant No. NSF PHY11-25915.

\bibliographystyle{apsrev4-1}
\bibliography{bibliography}
\end{document}